\documentclass[aps,prl,twocolumn,showpacs,superscriptaddress,floatfix]{revtex4}

\usepackage{graphicx}
\usepackage{epsfig}
\usepackage[english]{babel}

\newcommand{\be}{\begin{equation}}
\newcommand{\bea}{\begin{eqnarray}}
\newcommand{\eea}{\end{eqnarray}}
\newcommand{\ee}{\end{equation}}

\def\one{\ensuremath{\hbox{$\mathrm I$\kern-.6em$\mathrm 1$}}}
\begin{document}

\title{Image compression and entanglement }

\author{Jos\'e I. \surname{ Latorre}}
\affiliation{Dept. d'Estructura i Constituents de la Mat\`eria,
Univ. Barcelona, 08028, Barcelona, Spain.}

%\date{\today}

\begin{abstract}
The pixel values of an image can be casted into a
real ket of a Hilbert space using an appropriate 
 block structured addressing. 
The resulting state can then be
rewritten in terms of its  matrix product state representation
in such a way  that quantum entanglement corresponds to 
classical correlations between different coarse-grained textures.
A truncation of the MPS representation is tantamount to a
compression of the original image. The resulting algorithm can
be improved adding a discrete Fourier transform preprocessing
and  a further entropic lossless compression. 

\end{abstract}


\maketitle

Any technique designed to faithfully  handle many-qubit  quantum states
must retain as much entanglement as possible. This central
idea is present in all developments emerging from
the matrix product state representation of states \cite{Fannes}
and their generalization to projective entangled pairs \cite{PEPS}.
It is  clear that the very difficulty of 
handling entanglement on a classical computer is rooted
in the direct product structure of many-body Hilbert spaces.
Whatever is learnt about powerful representations
and manipulations of quantum states should readily translate
to any other classical problem with a large
direct product structure.

We here present an amusing proposal to compress
images using theoretical elements of quantum mechanics.
The algorithm works in three steps. 
We first cast an arbitrary 
image into a quantum register that only uses logarithmically
many quantum local degrees of freedom. The
entanglement of this state reflects the way
individual pixel values are correlated as a
due to their relative position in the image. 
We shall see that a renormalization group
inspired addressing of pixels suites the purpose
of writing the image as a real ket.
Second, we rewrite the resulting 
quantum state using the Matrix Product State (MPS) representation.
It will be seen that pictures with smooth textures
carry little entanglement and thus can be efficiently
represented using this construction. Third, we observe that
any truncation scheme for  entanglement entails 
a classical compression algorithm. 

The goal of the above algorithm is not to compete
(though it may be worth analyzing such a possibility)
with state-of-the-art  image compression techniques 
that take advantage of the detailed workings of human vision but,
rather, to 
explore the possibilities of spinning-off 
accumulated knowledge from quantum mechanics 
to classical problems with a direct product
structure.

{\sl 1. Casting an image into a real ket.}
Let us make our discussion concrete starting
with a 0 to 255 grey scale $2^n\times 2^n$ -pixel  image.
We can start to address each pixel using a  blocked structure
construction by taking an initial box,
labeled $i_1$, of $2\times 2$ pixels. So far, we have
only four pixels whose level of grey is defined
by numbers that we organize as
a ket in a real qudit, that is, a vector space of dimension 4,
\be
\label{block1}
|\psi_{2^1\times 2^1}\rangle= \sum_{i_1=1,\dots 4}c_{i_1}|i_1\rangle .
\ee
The value $i_1=1$ can be understood as labeling the up-left pixel, 
$i_1=2$ as the up-right one, $i_1=3$ as down-left one and 
$i_1=4$ as down-right one.
We now consider a larger block made of 4 inner sub-blocks as
shown in Fig.1.  
To identify which sub-block we
are addressing, a new label is needed, called $i_2$, with the
same convention as  defined for the inner block.
The new image displays a total of $2^2\times 2^2$ pixels and 
is represented by the real vector in $R^4\otimes R^4$
\be
\label{block2}
|\psi_{2^2\times 2^2}\rangle= \sum_{i_1,i_2=1,\dots 4}
c_{i_2,i_1}|i_2,i_1\rangle ,
\ee
where $c_{i_2,i_1}$ store all the pixel values.

\begin{figure}[t]
  %\centering
  \includegraphics[width=0.4\linewidth]{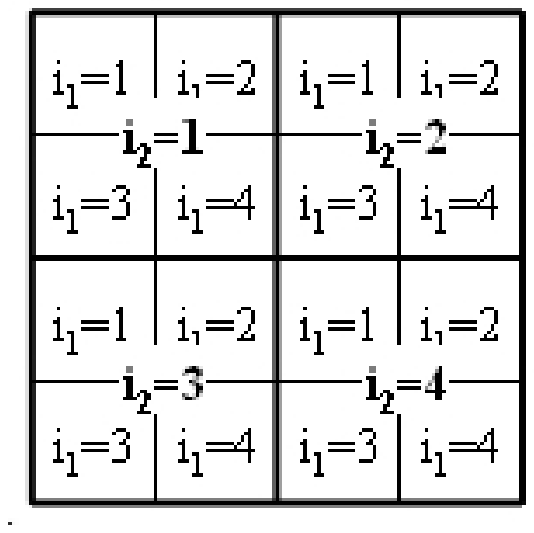}
  \caption{ Renormalization group inspired addressing of
pixel positions suited to cast an image into a
real ket. Each  qudit carries a partial information
of the color of a pixel at a different scale. The figure
exemplifies the way an image with a total of $4^2$ pixels is
casted into the state $
|\psi_{2^2\times 2^2}\rangle= \sum_{i1,i2=1,\dots 4}
c_{i_2,i_1}|i_2,i_1\rangle .
$ where pixel values are stored in $c_{i_2,i_1}$.
 }
  \label{FigEnergy}
\end{figure}

This block structure can be extended an arbitrary 
number of steps up to a size $2^{n}\times 2^{n}$. The representation of
the image corresponds to the real ket  in $\left(R^4\right)^{\otimes n}$
\be
\label{blockn}
|\psi_{2^{n}\times 2^{n}}\rangle= \sum_{i_1,...,i_n=1,\dots 4}
c_{i_n,\dots,i_1}|i_n,\dots,i_1\rangle .
\ee
At this point, an image with $2^n\times 2^n$ pixels is represented as 
a n-qudit real quantum state that we need not normalize.

Let us pause to reflect on some of the properties of the representation
we have just constructed. First, the number of qudits needed
grows logarithmically with the total size of the picture. This is possible
because the individual pixels are stored as coefficients of basis states,
addressed  in a telescopic, renormalization group manner. Each qudit is
in charge of retaining in which quadrant the pixel lives at a
given coarse grained  level. As a consequence,
the quantum state which represents
the image must be highly entangled. Generically, the state will be
maximally entangled for a random image. Nevertheless, this is not
the case of the images that we are used to see as they typically carry 
extended structures.
 A second observation can be
made about the meaning of this entanglement. Since every qudit is
attached to a different coarse graining level, entanglement between
adjacent qudits quantifies the increasing richness of textures
as we fine-grain further and further the image. The more surprising the
finer details of the image
are, the more independent superpositions are needed. On the other hand, 
smooth surfaces need less superposed states to represent them.  

Let us illustrate the proposed quantum encoding of images with some examples.
A plain white image is just a series of 255's as the grey scale value
of every pixel. This amounts to an equal superposition of
every basis state, which is a product state. No entanglement is needed
because no texture is carried by the picture. An image made with
four quadrants of different levels of grey will be represented 
by a superposition of just four states. The more complex the 
picture is, the more non-separable superpositions we shall find.
This means that zones with flat textures will need only little
entanglement between the qudits involved in determining that region.

{\sl 2. Matrix product representation of an image.}
The second step of the algorithm is to convert the representation of the image
as constructed above in the computational basis into a
matrix product representation \cite{Fannes}. 
The well-known idea is to find $n$ real
tensors $\Gamma^{(a)i_a}_{\alpha_a,\alpha_{a+1}}$,
$a=1,\dots,n$ with  physical indices $i_a=1,\dots,4$ and 
two ancillae indices $\alpha_a=1,\dots,\chi_a$ and
$\alpha_{a+1}=1,\dots,\chi_{a+1}$ 
so that
\be
\label{MPS}
|\psi_{2^{n}\times 2^{n}}\rangle= 
\sum_{i's}\sum_{\alpha's}\Gamma_{\alpha_1\alpha_2}^{(1)i_1}
\Gamma_{\alpha_2\alpha_3}^{(2)i_2}\dots \Gamma^{(n)i_n}_{\alpha_n\alpha_1}
|i_n,\dots i_2,i_1\rangle
\ee
where the register is treated as periodic for convenience. 
The usual interpretation of this representation applies here. Each
tensor can be viewed as a projector from a pair of ancillae to a physical
degree of freedom. The manifest advantage of MPS is that the range of
the ancillae space is related to the amount of entanglement between
qudits. 

The quantitative relation between entanglement and the range of ancillary
indices can be understood in a simple way by taking
the  range for the first ancilla index $\alpha_1$ to be just one. 
This choice eliminates {\sl de facto} the periodic boundary conditions, so that
we are considering a linear chain rather than a ring
of qudits to represent the original image. Such a representation can
be constructed operating a series of successive Schmidt decompositions
\cite{Vidal}. More concretely, the range $\chi_a$ of  index $\alpha_a$
corresponds to the Schmidt number of the partition of the state between
the first $(a-1)$ qudits and the rest. This in turn puts
a bound to the von Neumann entropy $2^{S_{a}}\le \chi_a\le
4^a$. Note
that the maximum possible $\chi_a\le \chi_{max}$ appears
at half of the chain and
corresponds to $\chi_{max}=4^{n/2}$.
We have now
a better understanding of the quantum representation of the classical
image. The more random the image, the more entropic the correlations
between coarse-graining levels are 
and the larger the range for the ancillae should be.

We immediately find a first result. Let's consider an 
made up image such
that its exact MPS representation carries little entanglement,
that is, the ranges $\chi_a$ are far less than their allowed maximum.
Such a picture would have dominant relations between blocks and would
definitely not look random. In such a case, the MPS quantum representation
of the image would be extremely efficient as compare to the
pixel based representation. Moreover, the gain obtained using the MPS
representation would be exponentially
large if ancillae indices only range up to a polynomial function
in $a$ rather than exponential one. This would imply that we
could store and send the exact content of an image using
the set $\{\Gamma^a\}$. This lossless compression could be named
{\sl qzip} in the sense that it would be exact
and that  it would saturate the von Neumann entropy associated
to adjacent two-party partitions in the register.

Let's note that {\sl qzip} is devised in a completely different
way to the entropic lossless {\sl gzip} compression algorithm
(and all other general purpose lossless zips based on the Lempel-Liv algorithm
\cite{LZ77}), 
which  saturates Shannon's entropy of the file
as given by a linear sequence of bits.
In general,   
{\sl gzip} will be vastly superior to {\sl qzip} unless a definite
block structure is present in the image. In this sense, {\sl qzip}
is just an academic observation. 
 Yet, it is readily checked
that a picture  described exactly by a reduced set of 
$\{\Gamma^{(a)}\}$ needs more data to be  kept when
expanded in pixels and then compressed with {\sl gzip}. 
The basic idea is that in such a case {\sl qzip} stores
the values of exponentially many bits as a product of polynomially-many 
matrices. The larger the block-structured picture, the more efficient
lossless compression using {\sl qzip} would be. Of course, standard
pictures are only partially block-structured and other
lossless algorithms are more efficient. This suggest to 
introduce our third step, that is, a truncation scheme for entanglement.

{\sl 3. ``Quantum'' compression of an image: qpeg.}
The MPS representation of a ket opens a road to define
truncation schemes that favor a {\sl bona fide} representation of
entanglement. The
idea is simple: we can truncate the ancillae space to our best
convenience. We could proceed in two ways. We could start
with the exact MPS reprentation and then only retain the
highest eigenvalues in each Schmidt decomposition up to
a maximum we can choose {\sl a priori} \cite{Vidal}. A different strategy
consists in finding the truncated state which minimizes its
distance to the original state \cite{PEPS}. We shall follow this second,
exact approach.

Basically, we want to approximate a state $|\psi(\Gamma)\rangle$ 
that codes the original image by $|\tilde\psi(\tilde\Gamma)\rangle$
that will code a compressed version of the same image,
where the range of ancillae indices $\alpha_a\le \chi_{trun}$ 
are truncated in $\tilde\Gamma^{(a)i_a}_{\alpha_a,\alpha_{a+1}}$. 
Therefore, the level of compression of {\sl qpeg} is defined by 
$\chi_{trunc}$, which must be far smaller than the allowed
maximum $\chi_{max}=4^{n/2}$.
The condition
of optimal compression based on a pixel-by-pixel criterion corresponds
to minimize the error function 
\be
\label{distance}
 \min_{\{\tilde\Gamma\} } \left| |\psi(\Gamma)\rangle -
| \tilde\psi(\tilde\Gamma)\rangle\right|^2 \ .
\ee
This expression is quadratic in the 
variables $\tilde\Gamma$. Elementary 
algebra leads to the system of equations
\be
\label{system}
\sum_{\alpha_a'\alpha_{a+1}'} \Gamma^{(a)i_a}
B_{(\alpha_a\alpha_{a+1})(\alpha_a'\alpha_{a+1}')}=
E^{i_a}_{(\alpha_a\alpha_{a+1})}
\ee
where all parenthesis indicate a combined index,
\be
\label{B}
B_{(\alpha_a\alpha_{a+1})(\alpha_a'\alpha_{a+1}')}\equiv
\sum_{i\not=i_a}
A^{i_1,\dots\hat{\imath}_a,\dots ,i_n}_{\alpha_a'\alpha_{a+1}'}
A^{i_1,\dots\hat{\imath}_a,\dots ,i_n}_{\alpha_a\alpha_{a+1}}\ ,
\ee
\be
\label{C}
E^{i_a}_{(\alpha_a\alpha_{a+1})}\equiv
\sum_{i\not=i_a}
c_{i_1,\dots,i_n} A^{i_1,\dots,\hat{\imath}_a,\dots,i_n}_{\alpha_a\alpha_{a+1}} \ ,
\ee
\be
\label{A}
A^{i_1,\dots\hat\imath_a,\dots ,i_n}_{\alpha_a\alpha_{a+1}}
\equiv \sum_{\alpha\not=\alpha_a,\alpha_{a+1}}
\Gamma^{(1)i_1}_{\alpha_1\alpha_2}\dots{\hat\Gamma}^{(a)i_a}_{\alpha_a\alpha_{a+1}}
\dots \Gamma^{(n)i_n}_{\alpha_n\alpha_1}
\ee
and where hat symbols are not present. This system is readily inverted
\be
\label{solution}
\Gamma^{(a)i_a}_{\alpha_a\alpha_a'}=(B^{-1})_{(\alpha_a\alpha_{a+1})
(\alpha_a'\alpha_{a+1}')}E^{i_a}_{(\alpha_a'\alpha_{a+1}')}\ .
\ee
The minimization algorithm must now proceed by sweeps of
the whole register. At every step, all $\Gamma$'s are kept
fixed but the one which is improved. The iterative procedure converges due
to the uniqueness of the minimum which is a consequence of 
the fact that we are Eq. (\ref{distance})  corresponds to a quadratic form
in $\tilde\Gamma $. 

The computing time to achieve a numerically acceptable minimum 
depends on the size of the register. Large images compressed to a large set of 
$\Gamma$'s will need a long minimization period. Simple and
small images will be delivered much faster. In the sequel we shall
work with a few local qunits, which makes compression
fast.

{\sl 4. Improvements.} Image compression can be substantially
improved taking advantage of some special features 
related to human vision. In particular, the elimination
of high frequency Fourier modes are often of no relevance  
to get a correct representation of an image (with some obvious exceptions like
astronomical pictures). This is actually used in popular compression formats
like {\sl jpeg} \cite{JPEG}. Such a compression algorithm 
first  discrete cosine Fourier transforms the initial image, 
applies a quantization matrix to get different accuracies for
each frequency and, finally, uses
an entropic lossless compression on a zig-zag reading of momentum modes. 

We can, therefore, improve our previous algorithm 
using a momentum space preprocessing and adding a final compression
of the set of $\{\tilde\Gamma\}$ maintaining the new conceptual
power of the compression algorithm. 
The complete
sequence of the algorithm, that we can named {\sl qpeg}, reads as follows:
\begin{enumerate}
\item Divide the original image in boxes.
\item Apply a discrete cosine Fourier transform to the box. 
\item Cast the Fourier transformed box using a RG inspired addressing 
into a ket, $|\psi\rangle$.
\item Represent the ket using MPS, $|\psi(\Gamma)\rangle$.
\item Truncate the ancillary indices to a preassigned maximum $\chi_{trunc}$
and use Eq. (\ref{B}-\ref{solution}),
$|\psi(\hat \Gamma)\rangle$.
\item Perform a lossless compression on the set of actual values of
the matrices, $gzip[\{\hat\Gamma\}]$.
\end{enumerate}
Note that, in momentum space, the RG inspired addressing of Fourier modes 
makes good sense. In the discrete Fourier tranformed box, low frequencies correspond
to the upper-left corner, whereas high frequencies are represented
by the lower-right part of the box. Slashed diagonals correspond
to similar frequencies as we move from vertical to horizontal modes.
The RG addressing is now clear. Higher frequencies are packed in a
coarse grained substructure whereas lower frequencies are 
assembled in the opposite corner. RG addressing is, thus, appropriate to
the momentum representation of an image.

\begin{figure}[t]
  %\centering
  \includegraphics[width=0.45\linewidth]{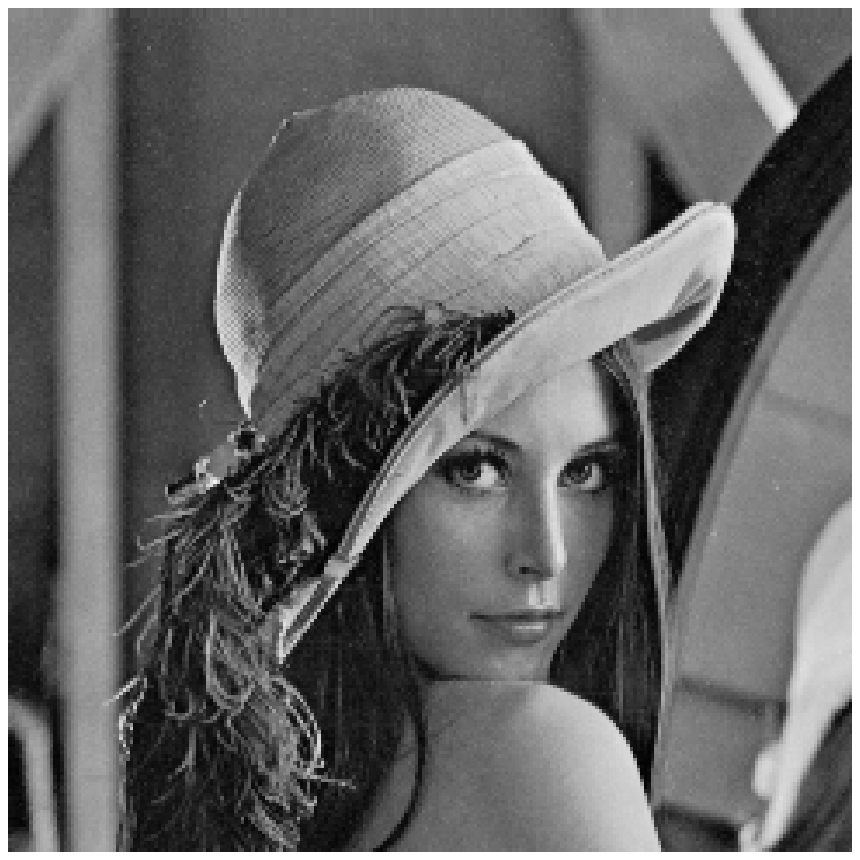}
  \includegraphics[width=0.45\linewidth]{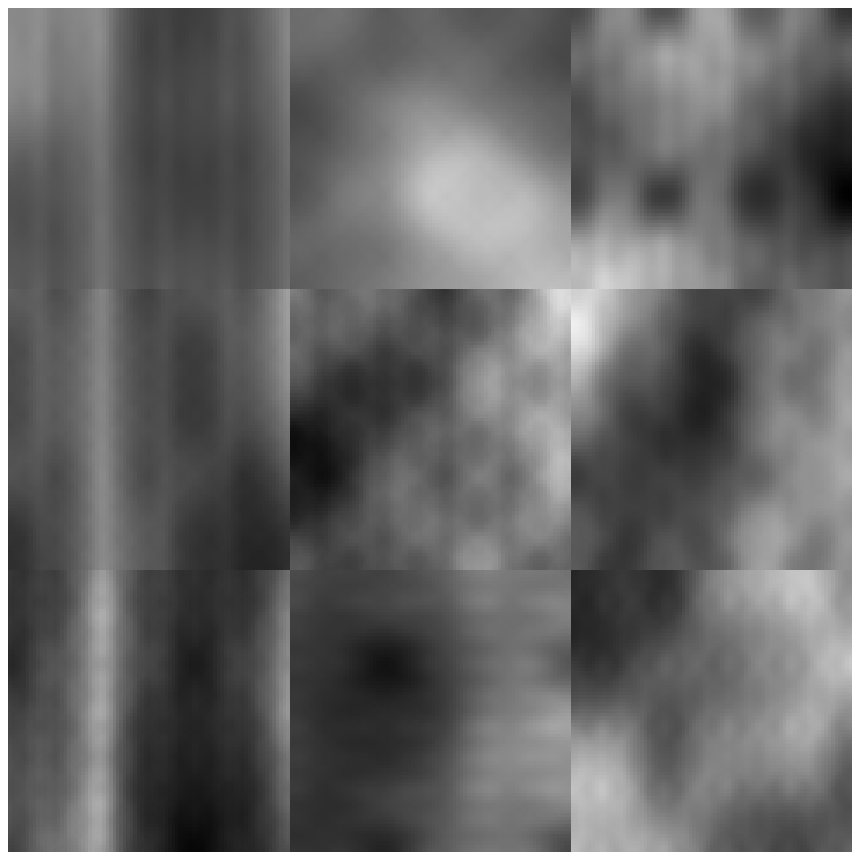}
  \includegraphics[width=0.45\linewidth]{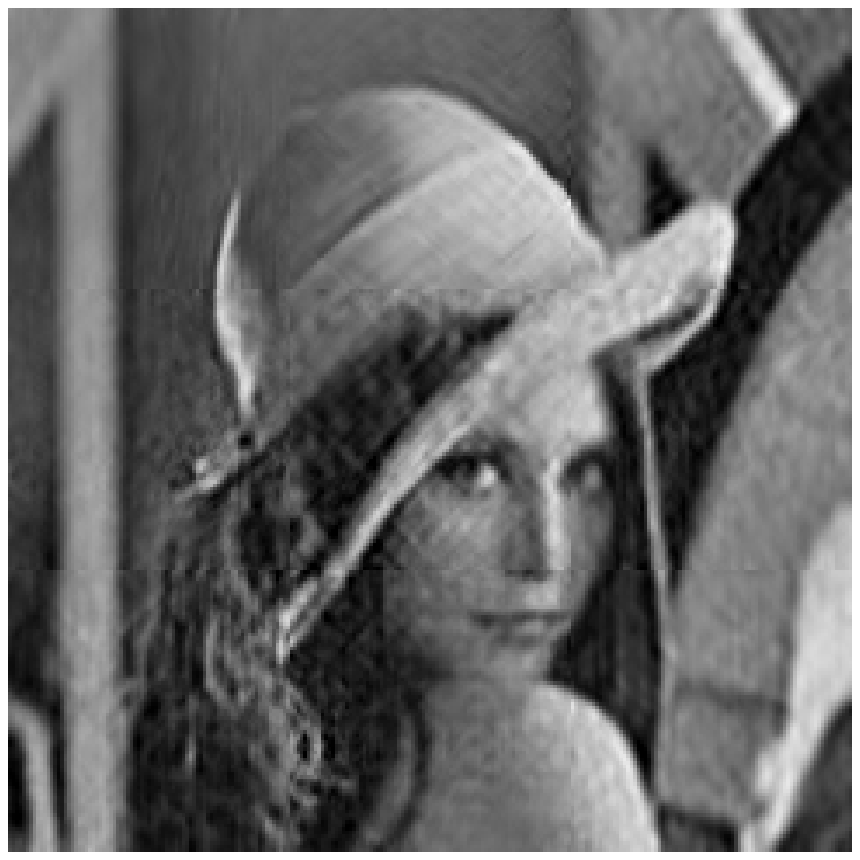}
  \includegraphics[width=0.45\linewidth]{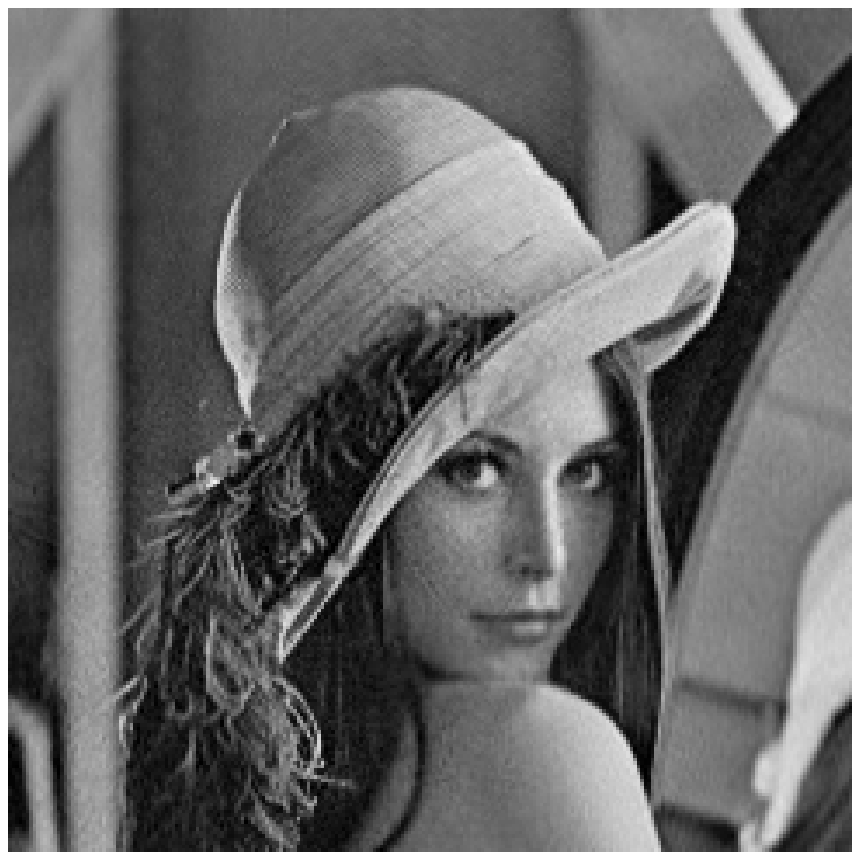}

  \caption{ The original 6561 pixels image in the upper-left corner is
compressed following the {\sl qpeg} algorithm with $\chi_{trunc}=1$
 (right-upper corner, 36 reals, PSNR=17), 
4 (left-lower corner, 576 reals, PSNR=25.6) and 8
 (right-lower corner, 2304 real, PSNR=31.9).
 }
  \label{FigImage}
\end{figure}

The  algorithm we have presented 
can be applied to a standard benchmark image as shown in
Fig. \ref{FigImage}. In this case, the image has been divided in 9 boxes
of $81\times 81$ pixels.
Each box is preprocessed separately with a discrete cosine 
Fourier transform. Then,
RG addressing of frequency-pixels builds up a real ket which is
further represented in terms of MPS. We have chosen for convenience
to use a register made of 4 qunits, each corresponding
to a 9 dimensional local Hilbert spaces (instead
of the qudits used in the general presentation). Different
truncations are, then, analyzed. The original size of each block contains
a total of 6561 integers that range between 0 and 255. 
The most dramatic truncation, shown in the upper-right corner of
Fig. \ref{FigImage}, only takes $\chi_{trunc}=1$, 
that is one-dimensional ancillae
for every qunit. Such a state carries no entanglement. 
Its representation in terms of $\Gamma^{(a)i_a}_{\alpha_a\alpha_{a+1}}$
needs only 4 matrices (index $a$) times 9 elements (index $i_a$)
since $\alpha_a=\alpha_{a+1}=1$, that is, a total of 36 real numbers
A common measure of the quality of compression is the PSNR measure in decibels
and defined
as $PSNR(dB)=10 \log_{10}(255^2/MSE)$, where $MSE=1/n \sum_{i}(y_i-x_i)^2$,
that is the sum over all $n$ pixels of the squared difference
between the original pixel value $x_i$ and the one resulting
from the compression $y_i$.  
In this case, PSNR=17.
A less severe truncation is shown in the left-lower corner of 
Fig.\ref{FigImage} with $\chi_{trunc}=4$ and $PSNR=25.6$.
The last compression shown corresponds to $\chi_{trunc}=8$ with
PSNR=31.9. It is desirable to achieve  PSNR larger than 30.

The ratio of stored bits per pixel is what really justifies a
good quality compression. In this sense further work
is needed to improve our basic scheme.
In particular it is possible to 
use adaptative dimensions for MPS and further approximate
the final values for the $\{\Gamma\}$ to a discrete predefined
series, such that its subsequent gzip compression would be more
efficient. Other preprocessing strategies, {\sl e.g.} wavelets or
preprocessing by quantization matrices (the selection of a good quantizer
appear to be instrumental to get a competitive algorithm), 
may also improved the global strategy.

The reader may wonder what the conceptual differences
between {\sl jpeg} and {\sl qpeg} are. As described above,
we have constructed {\sl qpeg} to use the same 
discrete Fourier preprocessing as {\sl jpeg} and both use
a final lossless entropic compression. The conceptual difference is
that {\sl qpeg} does not attempt to set to zero or to
approximate to a prescribed accuracy the set of frequencies
defining the image. Rather it tries to reproduce them as
best as possible as 
products of matrices, each one attached to a coarse-graining
level. The truncation in the indices of these matrices is
what makes {\sl qpeg} inexact. This hints at a possible
improvement of the basic algorithm based on allowing
the size of the matrices to locally adapt to the complexity of 
the texture of the image.

{\sl 5. Conclusion.}
Let us conclude with the general proposal that many 
complex classical problems are amenable to a quantum representation,
thus, allowing for the application of techniques
to handle entanglement. An obvious extension of the above
construction would be the casting of music files using a one-dimensional
RG addressing of frequencies to build a quantum state. More
dramatically,
information on three-dimensional objects is also easily compressed
by modifying the RG addressing and proceeding with the MPS representation and
truncation as stated above. 
It is also tantalizing to consider 
dynamics, that is, evolution of such a quantum representation
of an image. It might be arguable 
that all  information in an image could be stored
in a hamiltonian that would evolve an initial product state.

Financial support is acknowledge form MEC, QAP.
Part of this project was done at the Perimeter Institute.
It is a pleasure to thank discussions with  I. Cirac, S. Massar, R. Or\'us, 
 Ll. Torres and F. Vestraete.


\begin{thebibliography}{99}


\bibitem{Fannes} M. Fannes, B. Nachtergaele and R. F. Werner, Comm.
Math. Phys. {\bf 144}, 443 (1992).

\bibitem{PEPS} F. Verstraete and J.I. Cirac, cond-mat/0407066.

\bibitem{VPC04} F. Verstraete, D. Porras and J. I. Cirac, cond-mat/0404706.

\bibitem{Vidal}
G. Vidal, Phys. Rev. Lett. {\bf 91}, 147902 (2003);
G. Vidal, Phys. Rev. Lett. {\bf 93}, 040502 (2004).

\bibitem{LZ77}  Ziv J., Lempel A., ``A Universal Algorithm for Sequential Data
Compression,'' IEEE Transactions on Information Theory, Vol. 23, No. 3,
pp. 337-343.

\bibitem{JPEG} http://www.w3.org/Graphics/JPEG/

\end{thebibliography}
\end{document}